\documentclass[aps,prl,showpacs,notitlepage,nofootinbib,superscriptaddress,floatfix,showkeys,twocolumn]{revtex4-1}

\usepackage{blindtext}
\usepackage{hyperref}
\usepackage{amsmath,amssymb}
\usepackage{float}
\usepackage{microtype}
\usepackage{graphicx}
\usepackage{bm}
\usepackage{latexsym}
\usepackage{epsfig}
\usepackage{psfrag}
\usepackage{color}
\usepackage[dvipsnames]{xcolor}
\usepackage{subfigure}
\usepackage[section]{placeins}
\def\e1i{\epsilon_{1\mathrm{i}}}

\allowdisplaybreaks[1]

\begin{document}

\title{$\gamma$-ray and ultra-high energy neutrino background suppression due to solar radiation}

\author{Shyam~Balaji}
\email{sbalaji@lpthe.jussieu.fr}
\affiliation{Laboratoire de Physique Th\'{e}orique et Hautes Energies (LPTHE),\\
UMR 7589 CNRS \& Sorbonne Universit\'{e}, 4 Place Jussieu, F-75252, Paris, France}
\begin{abstract}
    The Sun emits copious amounts of photons and neutrinos in an approximately spatially isotropic distribution. Diffuse $\gamma$-rays and ultra-high energy (UHE) neutrinos from extragalactic sources may subsequently interact and annihilate with the emitted solar photons and neutrinos respectively. This will in turn induce an anisotropy in the cosmic ray (CR) background due to attenuation of the $\gamma$-ray and UHE neutrino flux by the solar radiation. Measuring this reduction, therefore, presents a simple and powerful astrophysical probe of electroweak interactions. In this letter we compute such anisotropies for TeV $\gamma$-rays, which at the Earth (Sun) can be at least $\simeq 10^{-4},(10^{-2})$. The optical depth at Earth for elongation angles more focused around the Sun ($\lesssim 10^\circ$), can be around $10^{-3}$ and larger. Neutrino attenuation is extremely tiny for for PeV scale UHE neutrinos. We briefly discuss observational prospects for experiments such as the Fermi Gamma-Ray Space Telescope Large Area Telescope (Fermi LAT), High-Altitude Water Cherenkov (HAWC) detector, The Large High Altitude Air Shower Observatory (LHAASO), Cherenkov Telescope Array (CTA) and IceCube. The potential for measuring $\gamma$-ray attenuation at orbital locations of other active satellites such as the Parker Solar Probe and James Webb Space Telescope (JWST) is also explored.
\end{abstract}
\maketitle

\section{Introduction}
The extragalactic background is a superposition of all radiation sources, both individual and diffuse, from the edge of the Milky Way to the edge of the observable universe, and is thus expected to encode a wide range of phenomena \cite{Lamastra:2017iyo,Finke:2022uvv}. It is broadly comprised of electromagnetic radiation and neutrinos which have a characteristic energy density and spectrum. Contributions are guaranteed from established extragalactic $\gamma$-ray source classes such as active-galactic-nuclei (AGN), star-forming galaxies, and $\gamma$-ray bursts \cite{Owen2021}. It provides a non-thermal perspective on the cosmos, which is also explored by the cosmic radio background, extragalactic cosmic rays (CRs), and neutrinos. 

 The electromagnetic component often referred to as the extragalactic background light (EBL) is the total integrated flux of all photon emission over cosmic time \cite{Cooray:2016jrk,Singh:2021yoi,Yan-kun:2022nmh}. This EBL is dominated energetically by thermal relic radiation from the last scattering surface observed as the cosmic microwave background (CMB). Different physical processes characterize the EBL in each waveband: starlight in the optical, thermal dust emission in the infrared, and X-ray emission from AGN \cite{Mattila:2019ybk}.  

Similarly, for neutrinos, the high energy frontier is expected to be especially rich, with neutrinos from baryonic accelerators ($\gamma$-ray bursts, AGN, etc.) extending up to about $\simeq100$ GeV\cite{Lunardini:2013iwa}. At even higher energies, the ultra-high energy (UHE) regime will exhibit cosmogenic neutrinos and may even reveal the presence of topological defects, which could emit neutrinos via a variety of energy loss channels \cite{Lunardini:2013iwa}. For sources far beyond the gamma ray horizon, neutrinos may be the only probe because, even at the highest energies, they propagate freely up to cosmological distances. The observation of  UHE particles such as photons, ions, and neutrinos provides critical information on astrophysical systems as well as the mechanisms of charged particle acceleration in these systems.

Since there is significant contamination from foregrounds, direct measurements of the EBL and neutrino background are difficult, necessitating the use of indirect observational or theoretical determinations to obtain an estimate of emitter populations \cite{caddy2022towards}. Exotic contributions, such as those resulting from a possible connection between dark matter and Standard Model (SM) particles, may be present in addition to astrophysical emissions. This possibility was recently considered in Ref.~\cite{Bernal:2022xyi}. Here the axion, a pseudo-Nambu-Goldstone boson
initially proposed to solve the strong-CP problem—and
axion-like particles (ALPs) were considered as a dark matter candidate. The ALP can decay into two photons based on the ALP-photon coupling and can contribute to the EBL. $\gamma$-ray
attenuation is sensitive to multi-eV
ALP decays through their contribution to the EBL \cite{Kalashev:2018bra,Korochkin:2019pzr,Korochkin:2019qpe}. Hence, additionally suppression of the $\gamma$-ray background due to the Sun as discussed in this work, could also affect constraints on dark matter candidates such as ALPs in the event of significantly improved flux resolution. 

The interaction between the CMB and the cosmic neutrino backgrounds with $\gamma$-rays and UHE neutrinos and the subsequent CR anisotropies have been studied in detail \cite{DOlivo:2005edp,Ruffini:2015oha,Franceschini:2021wkr,nikishov1961absorption,Fazio:1970pr}. Here, we will focus on a qualitatively different phenomena, computing instead the local anisotropy in the $\gamma$-ray and UHE neutrino background due to photon and neutrino emission from the Sun. A simple attempt at estimating the extinction of $\gamma$-rays due to sunlight was made in Ref. \cite{Loeb:2022pdm}. However the blackbody energy spectrum of the Sun was not included and the optical depth was estimated locally instead of being integrated over the entire region of the solar system where the sunlight and $\gamma$-rays are interacting. Additionally, the optical depth as a function of $\gamma$-ray energy or the observation angle from Earth relative to the Sun was not studied nor was the feasibility of probing the effect with $\gamma$-ray telescopes.

Here, we will include these effects and also consider the neutrino analogue whereby solar neutrinos annihilate with UHE antineutrinos or vice versa. We discuss the prospects for measuring the predicted $\gamma$-ray anisotropies with both space and ground-based experiments such as the Fermi Gamma-Ray Space Telescope Large Area Telescope (Fermi LAT), Cherenkov Telescope Array (CTA), the Large High Altitude Air Shower Observatory (LHAASO) as well as the High-Altitude Water Cherenkov (HAWC) experiment. We then discuss the possibility of measuring UHE neutrino anisotropies with the IceCube experiment. We also predict the expected attenuation at the Parker Solar Probe and James Webb Space Telescope (JWST) orbits and review the possibility of measuring EBL suppression due to the Sun at these locations in the future. Complete understanding of these anisotropies within the SM will enable constraints on beyond the SM models containing new states that could cause attenuation of the EBL or UHE neutrino background. In this letter, we will first outline the photon-photon and neutrino-antineutrino annihilation cross sections, then compute the optical depths and finally discuss experimental consequences. We will use natural units (where $\hbar=c=k_B=1$).


\section{Annihilation processes}

\subsection{Cross sections}
\begin{figure}
    \centering
    \includegraphics[width=0.4\textwidth]{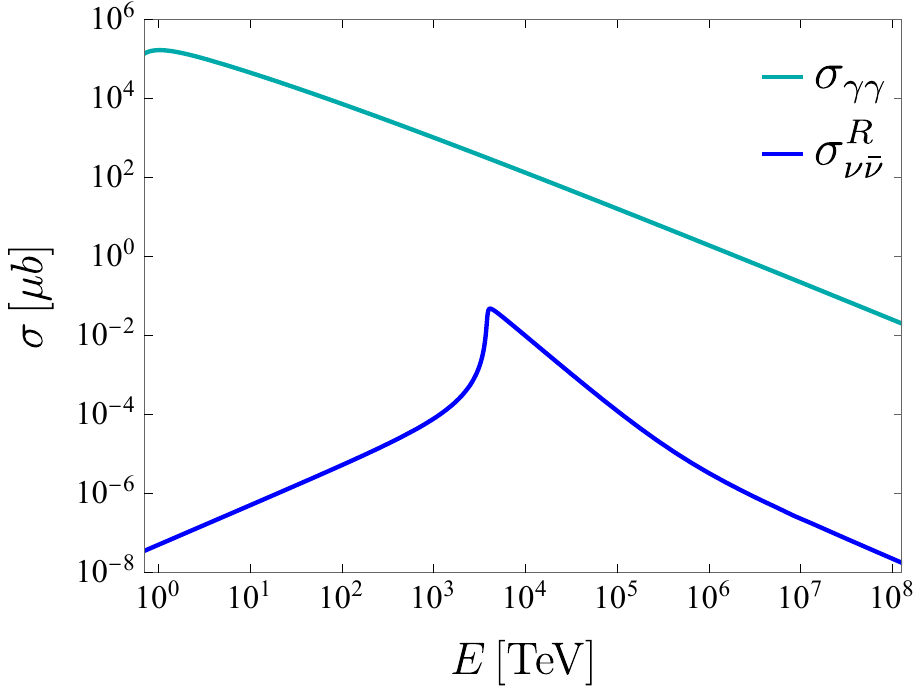}
    \includegraphics[width=0.4\textwidth]{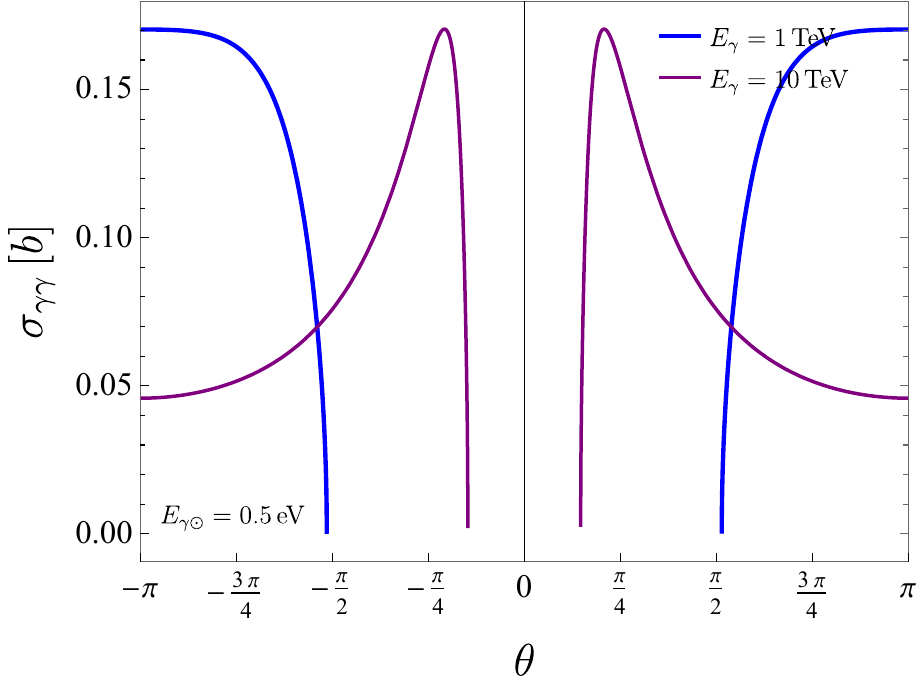}
    \caption{In the top panel we show annihilation cross sections in units of $\mu$b (microbarns) for $\gamma\gamma$ (cyan) and resonant $\nu\bar{\nu}$ (blue) as function of the incident $\gamma$-ray or ultra-high energy neutrino respectively. We take $E_{\gamma\odot}=0.5$eV and $E_{\nu\odot}=0.53$ MeV since these are typical energies for photons and neutrinos being emitted by the Sun. In the bottom panel we show the dependence of the annihilation cross section in b (barns) on the incident angle between the scattering photons for two energies $E_\gamma=1,\,10$ TeV while fixing $E_{\gamma\odot}=0.5$ eV.} 
    \label{fig:crosssection}
\end{figure}

    We will first consider the cross section for two photons annihilating into an electron-positron pair. For a $\gamma$-ray with energy $E_\gamma$, the annihilation cross section for the process $\gamma\gamma\rightarrow e^+ e^-$ \cite{Breit:1934zz,PhysRev.155.1404,Ruffini:2009hg} with a solar photon $\gamma_\odot$ of energy $E_{\gamma_\odot}$ is given by
\begin{align}
\label{eq:gammagammacrosssection}
    \sigma_{\gamma\gamma}(\beta) = &\frac{3\sigma_\textrm{T}}{16}(1-\beta^2)\times\nonumber\\&\times\left[2\beta(\beta^2-2)+(3-\beta^4)\log\left(\frac{1+\beta}{1-\beta}\right)\right],
\end{align}
where $\sigma_T=\frac{8\pi}{3}\left(\frac{\alpha}{m_e}\right)^2=6.652\times 10^{-25}\textrm{cm}^2$ is the Thomson scattering cross section, $\alpha$ is the fine structure constant, $m_e$ is the electron mass and a dimensionless kinematic factor is defined $\beta=\sqrt{1-E_\textrm{th}/E_\gamma}$. Where the threshold energy for electron-positron pair production is given by
\begin{align}
    E_\textrm{th}=\frac{2m_e^2}{E_{\gamma_\odot}(1-\cos\theta)}.
\end{align}
In this expression $\theta$ refers to the scattering angle between the $\gamma$-ray and the solar photon. The threshold energy of the incident $\gamma$-ray for this process to occur when annihilating with a photon of energy $E_{\gamma \odot}\simeq0.5$ eV (like the effective temperature of the Sun) is at least $\simeq 0.5$ TeV. Observing the bottom panel of Fig.~\ref{fig:crosssection}, we see that the scattering angle between the photons at which the cross section is maximised is highly dependent on the incident energy of the high energy $\gamma$-ray. Note that we will not consider higher order processes such as $\gamma\gamma\rightarrow e^+ e^- e^+ e^-$ since at high energy, the cross section approaches a relatively constant $6.5\,\mu b$ \cite{brown1973absorption} which is much lower than the leading order process unless $E_\gamma\gtrsim10^8$ TeV. Annihilation of photons into other final states such as $\mu^+ \mu^-$ or $\pi^+\pi^-$ are subdominant at the scales of interest in this work.

For the neutrino channel, the resonant neutrino-antineutrino annihilation into a fermion-antifermion, $\nu\bar{\nu}\rightarrow Z^0 \rightarrow f\bar{f}$ occurs via the $s$-channel. It has Breit-Wigner shape and can be written \cite{DOlivo:2005edp,Lunardini:2013iwa}
\begin{align}
\label{eq:nuresontant}
    \sigma_{\nu\bar{\nu}}^R(p,k)=\frac{G_F \Gamma m_Z}{2\sqrt{2} k^2 p E_{\nu\odot}}\int_{s_{-}}^{s_+} \frac{s(s-2m_\nu^2)}{(s-m_Z^2)^2+\xi s^2}ds,
\end{align}
where the Fermi constant is $G_F=1.16637\times 10^{-5} \textrm{GeV}^{-2}$ \cite{ParticleDataGroup:2020ssz}. The light neutrino mass is set to $m_\nu=0.08$eV and we define a dimensionless ratio $\xi=\Gamma^2/m_Z^2$, where $\Gamma=2.495$GeV is the decay width of the $Z$ boson with mass $m_Z=91.1876$GeV \cite{ParticleDataGroup:2020ssz}. The UHE neutrino has 4-momentum $k^\mu=(E_\nu,\mathbf{k})$ while the solar neutrinos have $p^\mu=(E_{\nu \odot},\mathbf{p})$, hence it follows from the usual relativistic energy-momentum relation that $k=\sqrt{E_{\nu}^2-m_\nu^2}$ and $p=\sqrt{E_{\nu\odot}^2-m_\nu^2}$. The centre-of-mass energy is $s=(p^\mu+k^\mu)^2\simeq2 m_\nu^2 + 2 \mathbf{k}\cdot (E_{\nu\odot} \pm \mathbf{p})$ since $E_\nu \simeq k$ at high energy. Also, we have $\mathbf{k}\cdot\mathbf{p}=p k\cos\theta$. Hence the integration limits in Eq.~\eqref{eq:nuresontant} are defined $s_{\pm}=2 m_\nu^2 + 2 k(E_{\nu\odot}\pm p)$ corresponding to $\theta=0$ and $\theta=\pi$ respectively.

We plot the resulting cross sections for photon-photon and resonant neutrino-antineutino annihilation as a function of the incident cosmic ray particle energy in Fig.~\ref{fig:crosssection}. We fix $E_{\gamma \odot}=0.5$ eV and $E_{\nu\odot}=0.53$ MeV respectively since these are the typical energies of photons and neutrinos being emitted from the Sun. The maximal cross section of $\sigma_{\gamma\gamma}\simeq1.7\times 10^5\,\mu$b for $\gamma\gamma$ occurs at $E_\gamma\simeq 1$ TeV  while for $\nu\bar{\nu}$ it occurs at $E_\nu\simeq 4\times 10^3$ TeV at $\sigma_{\nu\bar{\nu}}^R\simeq 4.8\times 10^{-2}\mu$b.

There are also non-resonant contributions, which include several other channels with final states such as $\nu\nu\rightarrow\nu\bar{\nu},l\bar{l}, WW, ZZ$ and $Zh$ \cite{Roulet:1992pz,Barenboim:2004di}, which we can approximate in total as in Ref.~\cite{Ruffini:2015oha} with
\begin{align}
    \sigma_{\nu\bar{\nu}}^{NR}\simeq \frac{\sigma^{he}_{\nu\bar{\nu}}}{1+E_r/E},
\end{align}
where $E_r=\frac{m_Z^2}{2m_\nu}$ and $\sigma^{he}_{\nu\bar{\nu}}=8.3\times 10^{-4}\, \mu b$. This is significantly smaller than the resonant contribution below $E_\textrm{th}$ so we can safely omit these contributions when considering UHE neutrino scattering in the regime of interest.

\subsection{Optical depth}
\label{sec:opticaldepths}
\begin{figure}
        \centering
    \includegraphics[width=0.4\textwidth]{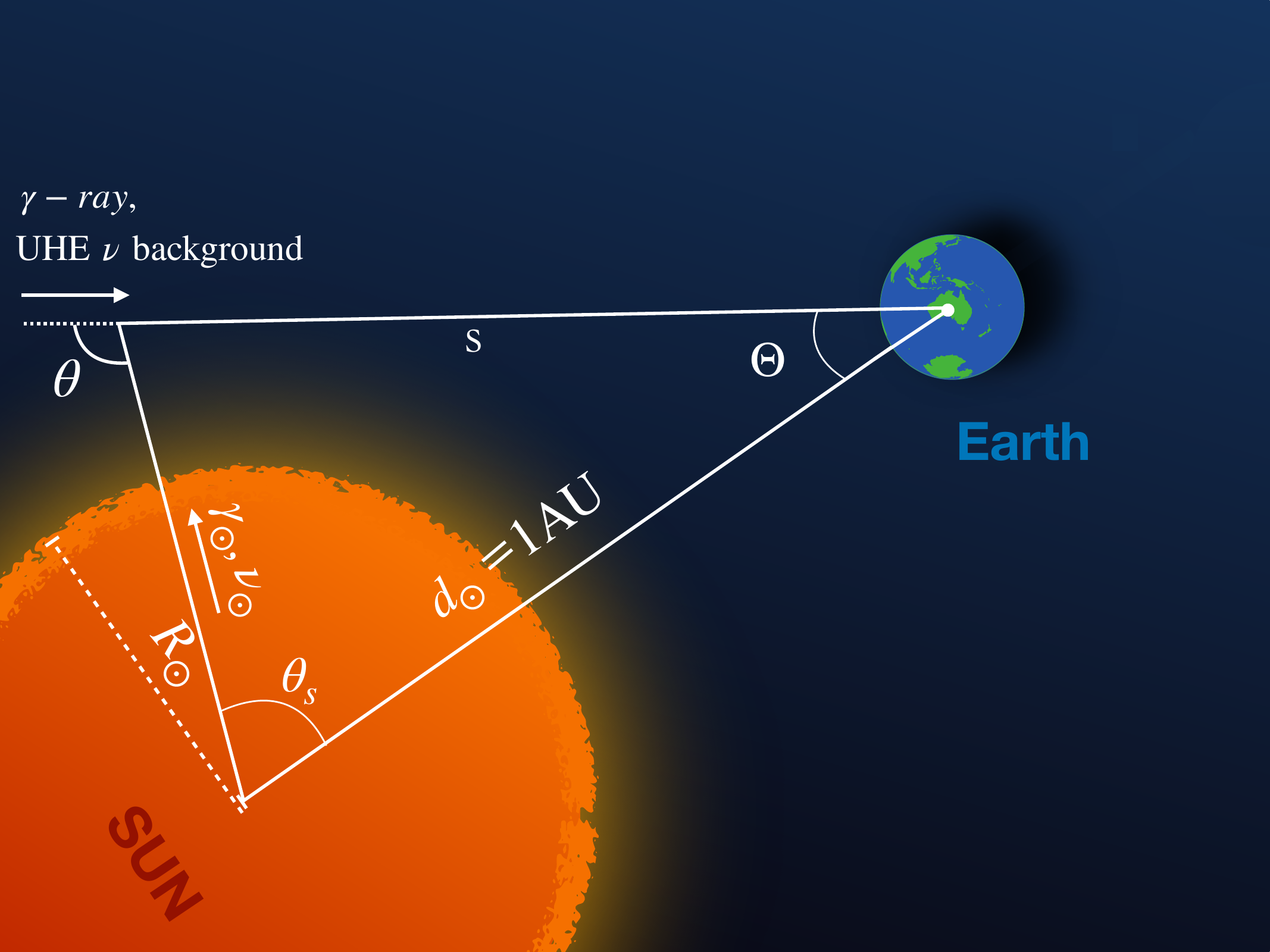}
    \caption{Representation of coordinate system showing line-of-sight distance $s$ where the extragalactic background photons and neutrinos interact with solar photons $\gamma_\odot$ and neutrinos $\nu_\odot$. The elongation angle $\Theta$, solar angle $\theta_S$ and scattering angle $\theta$ are also shown along with the solar radius $R_\odot$. }
    \label{fig:geometry}
\end{figure}

In natural units (with $\hslash=c=1$), the line-of-sight integrated optical depth as a function of incident $\gamma$-ray energy and elongation angle is given by \cite{Bottcher:2005pj} \footnote{The previous version of this manuscript was missing an angular factor as pointed out by Ref.~\cite{Finke:2024ada}, this has now been included below.}

\begin{equation}
\label{eq:opticaldepth}
\begin{split}
\tau_{\gamma}(E_\gamma,\Theta) = & \int_{0}^{\infty} ds \, (1-\cos\theta) \\
& \times \int_{E_{\text{th}}}^{\infty} dE_{\gamma_\odot} \, \sigma_{\gamma\gamma}(E_\gamma,E_{\gamma_\odot},\theta) \, n(s,E_{\gamma_\odot})
\end{split}
\end{equation}

where the radiation spectrum of the Sun can be approximated as a blackbody. The peak wavelength is around $500$ nm. The photon number density for the Sun modelled as a point source is given by
 \begin{align}
 \label{eq:photondensity}
n(s,E_{\gamma_\odot})=\frac{15}{4\pi^5}\frac{L_\odot E_{\gamma_\odot}^2}{T_\textrm{eff}^4(d_\odot^2+s^2-2d_\odot s \cos\Theta)(e^{E_{\gamma_\odot}/T_\textrm{eff}}-1)},
\end{align}
where $T_\textrm{eff}=5780$ K is the effective blackbody temperature of the Sun \cite{lide2007crc}, $L_\odot=4\times 10^{33}$ erg/s is the total luminosity of the Sun, $R_\odot=7\times 10^{10}$ cm is the solar radius and $d_\odot=1$ AU is the Earth-Sun distance. We can show a visual representation of the Sun-Earth-EBL(UHE $\nu$) system in Fig.~\ref{fig:geometry}. There are non-thermal sources of electromagnetic radiation from the Sun, such as X-rays and $\gamma$-rays from solar flares, X-rays and radio waves from coronal mass ejections, Sun spots and solar prominences. However, since the spectrum is dominated in intensity by sunlight, we will focus on this moving forward. In order to solve \eqref{eq:opticaldepth}, we require the scattering angle between background high energy photons and solar photons, $\theta$, as a function of $s$ and $\Theta$, this can simply be derived with trigonometric relations yielding
\begin{align}
\label{eq:cospsi}
    \cos\theta=\cos\theta_S\cos\Theta-\sqrt{1-\cos^2\theta_S}\sqrt{1-\cos^2\Theta},
\end{align}
where
\begin{align}
\label{eq:costhetas}
    \cos\theta_S=\frac{d_\odot-s\cos\Theta}{\sqrt{d_\odot^2+s^2-2d_\odot s\cos\Theta}}.
\end{align}
Hence, substituting \eqref{eq:costhetas} into \eqref{eq:cospsi} gives an exppression for $\theta$ as a function of $s$ and $\Theta$. Then the resulting expression for $\theta$ along with number density\eqref{eq:photondensity} can be substituted into \eqref{eq:opticaldepth} and integrated over line-of-sight distance $s$ and solar photon energy $E_{\gamma_\odot}$, enabling us to simply solve for the optical depth as a function of elongation angle $\Theta$ and incident $\gamma$-ray energy $E_\gamma$.

\begin{figure}
    \centering
    \includegraphics[width=0.45\textwidth ]{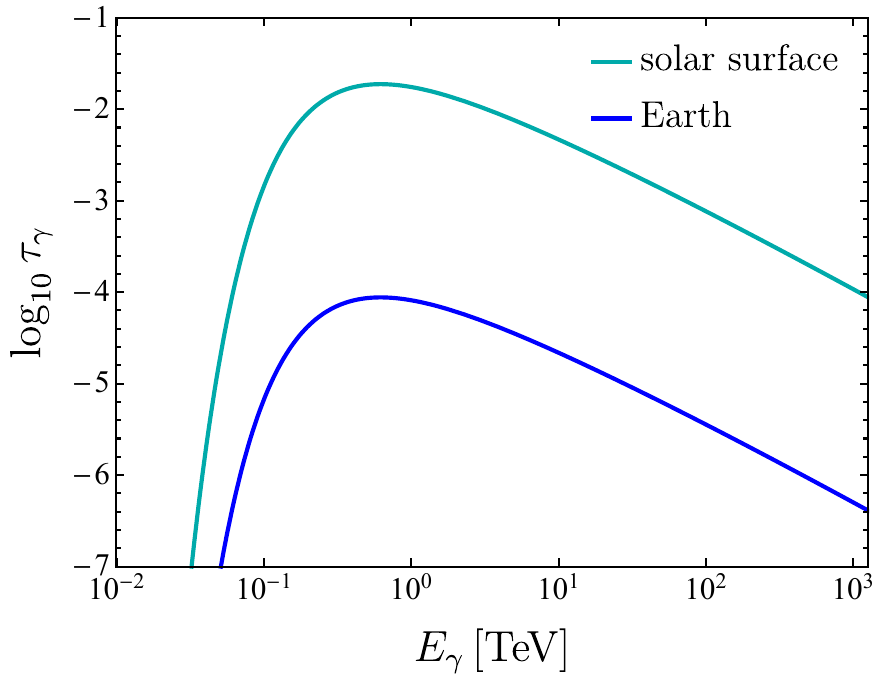}
            \includegraphics[width=0.45\textwidth ]{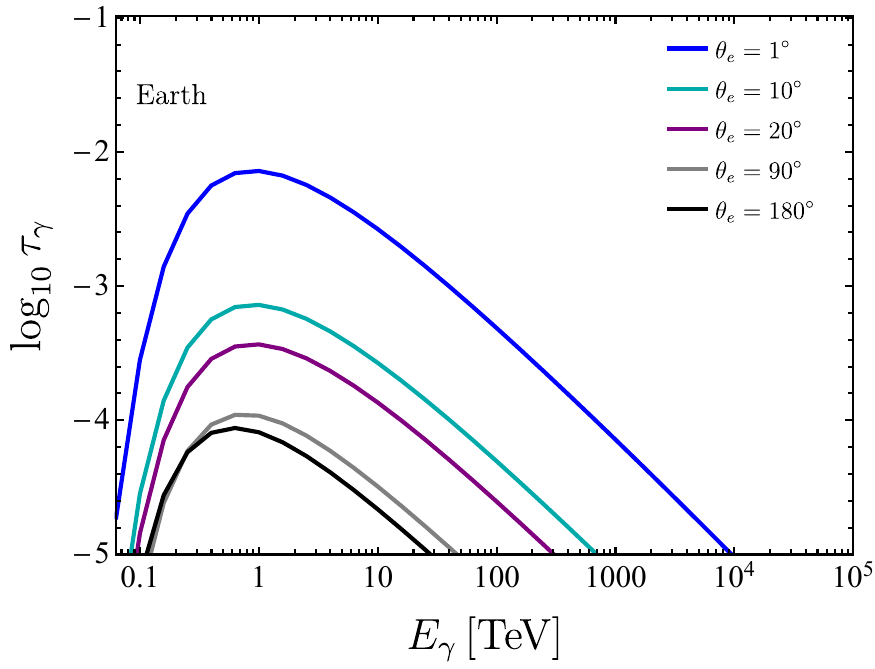}
        \includegraphics[width=0.45\textwidth ]{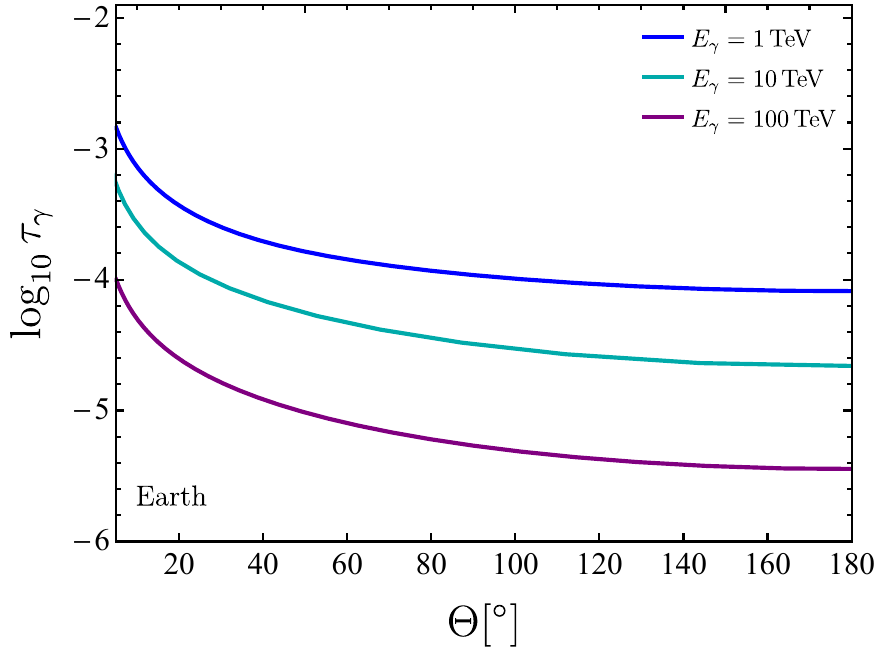}
    \caption{In the top panel we show the optical depth $\tau$ shown as a function of incident $\gamma$-ray energy in TeV. We show two scenarios, one is at the solar surface $r'=R_\odot$ (cyan) and the other is at Earth $r'=1$ AU for elongation angle $\Theta=\pm \pi$ (blue). In the middle panel, we plot the optical depth as a function of incident $\gamma$-ray energy as a function of elongation angle $\Theta$. In the bottom panel, we show optical depth at Earth as a function of elongation angle for three representative $\gamma$-ray energies, $E_\gamma=1$ TeV (cyan), $E_\gamma=10$ TeV (blue) and $E_\gamma=100$ TeV (purple).}
    \label{fig:gammarayopticaldepth}
\end{figure}

Hence we may easily obtain the dependence of the attenuation on the elongation angle $\Theta$, where $\Theta=0$ is in the direction  pointing directly towards the Sun from Earth (where the $\gamma$-ray background would be collinear with sunlight) and $\Theta=\pm \pi$ is pointing directly away from the Sun (where $\gamma$-rays would annihilate head on with sunlight). Note that $\theta$ is the scattering angle between incident photons, but $\Theta$ is the elongation angle at which observations from Earth would be performed. We may now plot the resulting optical depth due to the annihilation process as shown in Fig.~\ref{fig:gammarayopticaldepth} as a function of $E_\gamma$ for a fixed angle of $\theta=\pm \pi$ (head-on collision) at Earth and the solar surface, as shown in the top panel. In the middle panel, we show the attenuation as a function of incident $\gamma$-ray energy again, but this time, only at Earth and for several instructive elongation angles. Finally, in the bottom panel, we show the optical depth as a function of elongation angle relative to the Sun $\Theta$ for three representative energies $E_\gamma=[1,10,100]$ TeV in the bottom panel. 

For observational purposes, (i) lower $E_\gamma$ values and (ii) larger angles of $|\Theta|$ (pointing away from the Sun) are more relevant since the (i) flux uncertainties associated with $\gamma$-rays in this regime are lower and (ii) TeV photons from the solar corona at small $\Theta$ would be a very large background, potentially obscuring the observation of $\gamma$-ray attenuation. For this reason, we will conservatively focus our discussion on attenuation for $E_\gamma$ around a TeV at $\Theta=\pm \pi$. We also note that at small $|\Theta|$, there will be a large component of $\gamma$-rays resulting from a spatially extended component of the solar emission due to the inverse Compton scattering of CR electrons off solar photons \cite{Fermi-LAT:2011nwz}, this would make small effects like the one considered here difficult to resolve. However, it should also be noted that such a contribution is focused within a few degrees of the Sun and becomes insignifcant for elongation angles $|\Theta|\gtrsim 20^\circ$ \cite{Fermi-LAT:2011nwz}.  We show the optical depth at the solar surface and at Earth (for $\Theta=\pm \pi$) in cyan and blue in the upper panel of Fig.~\ref{fig:gammarayopticaldepth} respectively.  The maximal optical depth is for incident $\gamma$-ray energies around a TeV. In the first case, the $\gamma$-ray photons must traverse through the sunlight till the surface of the Sun and in the second, the $\gamma$-ray must travel through a much lower number density of solar photons from beyond the Earth and Sun towards the Earth's surface. Unsurprisingly, of the two scenarios shown, the optical depth is maximised at the solar surface at around $\tau_\gamma\simeq10^{-2}$ since the $\gamma$-rays interact with a much larger number density of surface photons emitted from the Sun. At Earth, the optical depth is around $\tau_\gamma=9\times10^{-5}$ for $E_\gamma=0.5$ TeV at angles of $\Theta=\pm \pi$ as shown in the top panel of Fig.~\ref{fig:gammarayopticaldepth}. In Ref.~\cite{Loeb:2022pdm}, the optical depth is estimated at $0.07$ at the solar surface and $3.3\times 10^{-4}$ at the Earth. In both cases, the estimate is noticeably different than what we obtain here. This is unsurprising since the energy distribution and radial dependence of the solar photons was not included or integrated over like here. 

We note that for elongation angles more focused around the Sun, the optical depth as observed from Earth is unsurprisingly larger than for $\Theta=\pi$. From the middle panel of Fig.~\ref{fig:gammarayopticaldepth}, we see that for $\Theta=1^\circ,10^\circ,20^\circ,90^\circ$ we get a max $\tau_\gamma\simeq7\times10^{-3},7\times10^{-4},4\times10^{-4},1\times10^{-4}$ respectively. This is simply due to the larger density of solar photons the EBL $\gamma$-rays can interact with while travelling to Earth. In the bottom panel of Fig.~\ref{fig:gammarayopticaldepth}, we see the optical depth for representative $\gamma$-ray energies plotted for elongation angles of $[5^\circ,180^\circ]$. We see that $\tau_\gamma$ increases as the angle $\Theta$ is reduced. There is a factor of a few more attenuation across all angles for $E_\gamma=1$ TeV relative to $E_\gamma=10$ TeV and a factor of more than five attenuation for $E_\gamma=10$ TeV relative to $E_\gamma=100$ TeV.

For UHE neutrino annihilation, we first require the solar neutrino luminosity which can be approximated $L_{\nu \odot}=0.023 L_\odot $ where $L_\odot\simeq 4\times 10^{33}$ erg/s is the total luminosity of the Sun \cite{Ianni:2014bqa}. We can approximate the neutrino number density at a distance $r$ with
\begin{align}
    n_{\nu\odot}(r) = \frac{1}{4\pi r^2} \frac{L_{\nu\odot}}{\bar{E}_{\nu\odot}},
\end{align}
taking $\bar{E}_{\nu\odot}=0.53$ MeV as the average neutrino energy emitted by the Sun \cite{Ianni:2014bqa}. Now we may compute the approximate optical depth $\tau_\nu$, for a high energy neutrino incident upon a solar neutrino
\begin{align}
    \tau_\nu = \int_{r'}^\infty n_{\nu\odot}(r) 
    \sigma_{\nu\bar{\nu}} dr,
\end{align}
using the maximum resonant neutrino annihilation cross section (at 4 PeV from Fig.~\ref{fig:crosssection} and Eq.~\eqref{eq:nuresontant}), we get  $\tau_\nu=1\times10^{-18}$ at Earth and $2\times 10^{-16}$ at the solar surface.

\section{Results and Discussion}
Considering the Fermi LAT experiment \cite{Fermi-LAT:2014ryh}, we see that the highest energy bin is $580$-$820$GeV. The reported intensity for this bin is $\simeq (9.7\pm 6.0 )\times 10^{-12}\,\textrm{cm}^{-2}\textrm{s}^{-1}\textrm{sr}^{-1}$. If we consider that the intensity will be similar at $1$ TeV, then we may multiply by the operating time of 1239 days, the angular coverage of 2.4 sr and an effective area of 8000 $\textrm{cm}^2$ \cite{Fermi-LAT:2014ryh} to get $20$ $\gamma$-ray events. Multiplied with the optical depth at Earth from Section \ref{sec:opticaldepths}, we get a reduction of $\simeq 2\times 10^{-3}$ events. This is substantially below unity,  so unless there is a factor of $\simeq 10^{3}$ improvement in event count, its unlikely that Fermi LAT will be able to register any reduction in events. Furthermore, Fermi LAT has a flux uncertainty of $\gtrsim 60\%$ for $\gamma$-rays at such high energy this makes small effects of attenuation difficult to discern.
The CTA is expected to become the largest and most sensitive observatory for very-high-energy $\gamma$-rays in the energy range from 20 GeV to more than 300 TeV. CTA will be capable of detecting $\gamma$-rays from extremely faint sources with unprecedented precision energy and angular resolution. CTA South is projected to have a peak differential energy flux sensitivity of around $E_\gamma^2 \frac{d\Phi}{d E_\gamma}\simeq1 0^{-13}\,\textrm{erg}\,\textrm{cm}^{-2}\textrm{s}^{-1}$ with an energy resolution of about 5\% and angular resolution of about $0.05^\circ$ for 1 TeV photons \cite{Maier:2019afm}. For $E_\gamma=1$ TeV and over a $4\pi$ angle, this corresponds to a differential flux sensitivity of $\simeq 5\times 10^{-15}\,\textrm{cm}^{-2}\,\textrm{s}^{-1}\,\textrm{sr}^{-1}$. This is $\gtrsim 3$ orders of magnitude better flux resolution than Fermi LAT. This suggests that one could hope for $\simeq 3\times 10^{-4} $ $\gamma$-ray flux resolution at TeV scale. This would perhaps be the most promising way to test attenuation of $\gamma$-rays due to sunlight, which as we calculated earlier, is of the order of $10^{-4}$ at Earth. Therefore, CTA seems a hopeful candidate for achieving such a goal, however in this case, there could be systematic uncertainties associated with resolving such a tiny effect. This is because Imaging Air Cherenkov Telescopes (IACTs) like CTA have a narrow field of view and local systematics like atmospheric conditions.  By combining different subsets of telescopes and pointing directions, CTA will be able to cover a large fraction of the Galactic plane with varying sensitivity and resolution. This could be effective for studying $\gamma$-ray attenuation at elongation angles away from the Sun. But, CTA will still be limited at probing elongation angles close to the Sun, as it will only record data at night.

The High Altitude Water Cherenkov Observatory (HAWC), completed in early 2015, has been used to observe the Crab Nebula at high
significance across nearly the full spectrum of energies to which HAWC is sensitive. HAWC’s sensitivity improves with
the $\gamma$-ray energy. For $E_\gamma\gtrsim1$ TeV the sensitivity is driven by the best background rejection and angular resolution
ever achieved for a wide-field ground array. The total uncertainty at 1 TeV for HAWC in this measurement according to Ref.~\cite{Abeysekara:2017mjj} appears to be around $\gtrsim 50$\% in total flux at 1 TeV. In the more recent Crab Nebula measurement Ref.~\cite{HAWC:2019xhp} by HAWC, the flux uncertainty is significantly improved at TeV scale, the flux is measured to be $(3.73\pm 0.07)\times 10^{-11}$ and $(3.63\pm 0.08)\times 10^{-11}\, \textrm{TeV}\,\textrm{cm}^{-2}\,\textrm{s}^{-1}$ respectively, depending on the energy resolution method used. This corresponds to a remarkable flux uncertainty of only $\simeq 2$\%.

LHAASO, can measure the energy and arrival direction of the $\gamma$-rays with high sensitivity and large effective area, and use them to probe the properties of the EBL, and thereby the solar attenuation effect described in this work. LHAASO has a larger effective area and a higher sensitivity for very-high-energy $\gamma$-rays than HAWC, especially in the multi-TeV and PeV range \cite{LHAASO:2023gne}. However, in both air shower detectors HAWC and LHAASO \cite{LHAASO:2023rpg}, their comparable flux resolution must improve significantly to probe solar attenuation of $\gamma$-rays at TeV scale.

More recently, an all-sky measurement of the anisotropy induced by cosmic rays travelling through our local interstellar medium and the interaction between the interstellar and heliospheric magnetic fields was performed \cite{HAWC:2018wju}. The analysis was based on data collected by the HAWC and IceCube observatories in the northern and southern hemispheres at the same median primary particle energy of 10 TeV. In their sky maps, they determined the horizontal anisotropies to be $\simeq 10^{-4}$. 
The vertical direction of the anisotropy was also determined to be of the same order. 
However it is important to note that this is for all cosmic rays, including charged cosmic rays, not only $\gamma$-rays like we are interested here. In Ref.~\cite{Abeysekara:2018qho}, they report an anisotropy of $8\times 10^{-4}$ at 2 TeV and  to $14\times 10^{-4}$ at around 30 TeV, from cosmic rays once again. They state that TeV cosmic-ray anisotropy
is primarily dipolar with amplitude $\simeq 10^{-3} $, but also contains smaller scale structure with strength $10^{-4}$. Improvements can be made with larger instantaneous sky coverage and longer uninterrupted observation periods. This would enable reduction in statistical uncertainties below the signal strength and to resolve features with large angular extent. To probe the solar attenuation of $\gamma$-rays, it is necessary to obtain anisotropy maps and flux resolution of $\gamma$-rays at the same level of sensitivity as for cosmic rays described above. This would require significant improvement in experimental design and analysis. Such improvements in sensitivity would be a significant step forward in understanding the effects of sunlight on $\gamma$-ray propagation.

In Ref.~\cite{Loeb:2022pdm}, $\gamma$-ray attenutation is compared with the dipole anisotropy measured by the Cosmic Background Explorer (COBE). This variation of $\simeq 1.2\times 10^{-3}$ results from the Earth's orbital motion about the Solar system barycenter  and strikingly lies between the EBL optical depths evaluated at the solar surface and Earth respectively \cite{Kogut:1993ag}. However, its also important to note that the CMB dipole anisotropy is at a significantly different energy scale and is not impacted by the same statistical and systematic uncertainties of the $\gamma$-ray sky at TeV scale.

The radial position of the Parker Solar Probe which orbits very close to the Sun's corona is around $9.86R_\odot$ from the solar centre~\cite{Wiedenbeck:2017ppg}. Supposing EBL measurements could be performed in this orbit, excellent extraction of $\gamma$-ray suppression could be discerned. In the case of an orbit with similar radial position to the JWST, which is situated at the L2 Lagrange point, we would get a smaller EBL optical depth $\tau_\gamma\simeq 10^{-4}$ at $E_\gamma=1$ TeV at $\Theta=\pm \pi$. This is almost the same as on Earth, but if $\gamma$-rays are recorded at L2, the Earth's umbra would not interfere with the sunlight $\gamma$-ray interactions, which could provide a marginal improvement in the measurement.

For UHE neutrinos, in order to obtain optimistically large optical depths of $\tau_\nu=1\times10^{-18}$ at Earth and $\tau_\nu=2\times 10^{-16}$ at the solar surface, we require energies of $E_\nu=4$ PeV. It is unlikely that such anisotropies will be probed with experiments such as IceCube which only have $\mathcal{O}(1)$ event observations at PeV scale which is far too small to study such an anisotropy \cite{IceCube:2018fhm}.
\section{Conclusion}
We consider suppression of diffuse extragalactic $\gamma$-rays and ultra-high-energy (UHE) neutrinos due to annihilation upon interactions with large numbers of photons and neutrinos emitted locally by the Sun. The annihilation induces anisotropies in the extragalactic background light (EBL) at around a TeV with optical depths of at least $\tau_\gamma \simeq 10^{-4}$ and $\tau_\gamma \simeq 10^{-2}$ at Earth and the surface of the Sun respectively. The optical depth at Earth can exceed $10^{-3}$ for smaller elongation angles. Such anisotropies are a direct prediction of Quantum Electrodynamics interactions and can only be probed with future $\gamma$-ray experiments such as Fermi LAT, HAWC, LHAASO and CTA, only if there is a significant improvement in the ability to measure flux and anisotropy resolution as well as reduction in systematic uncertainties. 

HAWC has measured dipole anisotropies of $\simeq 10^{-4}$ due to cosmic rays travelling through the interstellar medium. We obtain an optical depth of $\simeq2\times 10^{-5}$  at Earth with elongation angle $\pm \pi$. At energies much higher, the annihilation cross section falls off appreciably as does the EBL flux. Measurements of the TeV EBL suppression at comparable elongation angles can be larger around the Parker Solar Probe orbit $\simeq 10^{-3}$, while for telescopes at L2 such as the JWST, the anisotropy is smaller around $10^{-4}$. New experiments at these locations in the Solar System with $\gamma$-ray sensitivity could also probe EBL reduction. Experiments like HAWC and LHAASO would need to obtain sensitivities comparable to cosmic rays for the much smaller subset of $\gamma$-rays for this effect to be resolved. On the otherhand, CTA is expected to obtain flux resolution better than $0.1$\% for TeV $\gamma$-rays. This makes it an optimistic candidate to test the attenuation described in this work based on flux resolution, at elongation angles away from the Sun. However, even in this case, systematic uncertainties would need to be controlled to make the effect discernable. Hence we conclude that the comparisons in Ref.~\cite{Loeb:2022pdm} of the $\gamma$-ray anisotropy with the CMB dipole anisotropy are overly idealized given the much greater experimental challenges associated with resolving the EBL at TeV scale described in the main body of this work.

In the case of ultra-high energy (UHE) neutrinos interacting with the solar neutrinos of average energy $0.53$ MeV, we get much smaller optical depths. This is due to the smallness of the resonant and non-resonant contributions to the neutrino-antineutrino annihilation cross section. Even the most optimistic scenario can produce anisotropies of $\simeq 10^{-16}$ and $\simeq 10^{-18}$ for a PeV scale UHE neutrino scattering off 0.53 MeV solar neutrinos at the solar surface and at Earth respectively. Since experiments such as IceCube have only obtained a few PeV neutrino events, probing such a small anisotropy seems highly unrealistic with current experiments. 

The result for diffuse $\gamma$-rays presents a theoretical opportunity to study fundamental photon-photon interactions between isotropic background photons and thermal photons produced by the Sun in our neighbourhood of the universe. Although it seems unrealistic to probe such an attenuation with current telescopes, this result is an important SM effect and can be amplified with beyond the SM contributions in future works. We do not study the effects on non-diffuse background contributions in detail, but the calculated optical depths may easily be applied to these scenarios as well. 

\color{black}
\section{Acknowledgements}
 SB would like to thank Maura E. Ramirez-Quezada, Yongchao Zhang and an anonymous referee for helpful discussions and feedback on the draft. SB also thanks Justin D. Finke for pointing out a missing factor in the previous revision, which has now been included in this version to obtain the correct angular dependence for the optical depth. SB is supported by funding from the European Union’s Horizon 2020 research and innovation programme under grant agreement No.~101002846 (ERC CoG ``CosmoChart'') as well as support from the Initiative Physique des Infinis (IPI), a research training program of the Idex SUPER at Sorbonne Universit\'{e}. 

\section{Appendix}
The closed form solution for the resonant neutrino-antineutrino annihilation cross section in Eq.\eqref{eq:nuresontant} is given by
\begin{align}
    \sigma_{\nu\bar{\nu}}^R&(p,k)=\frac{2\sqrt{2}G_F\Gamma m_Z}{2 k E_{\nu \odot}}\left\{\frac{1}{1+\xi}+\frac{m_Z^2}{4 pk(1+\xi)^2}\log\left(\frac{f_+}{f_-}\right)\right.\nonumber\\
    &+\left.\frac{1-\xi}{(1+\xi)^2}\frac{m_Z^3}{4pk\Gamma}\left[\tan^{-1}(g_+)-\tan^{-1}(g_-)\right]\right\}
\end{align}
where $\xi=\Gamma^2/m_Z^2$ and
\begin{align}
    f_{\pm}&=4k^2(1+\xi)(E_{\nu\odot}\pm p)^2-4 m_Z^2 k (E_{\nu\odot}\pm p)+m_Z^4 \nonumber \\
    g_\pm&=\frac{2k(1+\xi)(E_{\nu\odot}\pm p)-m_Z^2}{\Gamma m_Z}
\end{align}
\bibliographystyle{apsrev4-1}
\bibliography{references.bib}

\begin{thebibliography}{42}%
\makeatletter
\providecommand \@ifxundefined [1]{%
 \@ifx{#1\undefined}
}%
\providecommand \@ifnum [1]{%
 \ifnum #1\expandafter \@firstoftwo
 \else \expandafter \@secondoftwo
 \fi
}%
\providecommand \@ifx [1]{%
 \ifx #1\expandafter \@firstoftwo
 \else \expandafter \@secondoftwo
 \fi
}%
\providecommand \natexlab [1]{#1}%
\providecommand \enquote  [1]{``#1''}%
\providecommand \bibnamefont  [1]{#1}%
\providecommand \bibfnamefont [1]{#1}%
\providecommand \citenamefont [1]{#1}%
\providecommand \href@noop [0]{\@secondoftwo}%
\providecommand \href [0]{\begingroup \@sanitize@url \@href}%
\providecommand \@href[1]{\@@startlink{#1}\@@href}%
\providecommand \@@href[1]{\endgroup#1\@@endlink}%
\providecommand \@sanitize@url [0]{\catcode `\\12\catcode `\$12\catcode
  `\&12\catcode `\#12\catcode `\^12\catcode `\_12\catcode `\%12\relax}%
\providecommand \@@startlink[1]{}%
\providecommand \@@endlink[0]{}%
\providecommand \url  [0]{\begingroup\@sanitize@url \@url }%
\providecommand \@url [1]{\endgroup\@href {#1}{\urlprefix }}%
\providecommand \urlprefix  [0]{URL }%
\providecommand \Eprint [0]{\href }%
\providecommand \doibase [0]{http://dx.doi.org/}%
\providecommand \selectlanguage [0]{\@gobble}%
\providecommand \bibinfo  [0]{\@secondoftwo}%
\providecommand \bibfield  [0]{\@secondoftwo}%
\providecommand \translation [1]{[#1]}%
\providecommand \BibitemOpen [0]{}%
\providecommand \bibitemStop [0]{}%
\providecommand \bibitemNoStop [0]{.\EOS\space}%
\providecommand \EOS [0]{\spacefactor3000\relax}%
\providecommand \BibitemShut  [1]{\csname bibitem#1\endcsname}%
\let\auto@bib@innerbib\@empty
\bibitem [{\citenamefont {Lamastra}\ \emph {et~al.}(2017)\citenamefont
  {Lamastra}, \citenamefont {Menci}, \citenamefont {Fiore}, \citenamefont
  {Antonelli}, \citenamefont {Colafrancesco}, \citenamefont {Guetta},\ and\
  \citenamefont {Stamerra}}]{Lamastra:2017iyo}%
  \BibitemOpen
  \bibfield  {author} {\bibinfo {author} {\bibfnamefont {A.}~\bibnamefont
  {Lamastra}}, \bibinfo {author} {\bibfnamefont {N.}~\bibnamefont {Menci}},
  \bibinfo {author} {\bibfnamefont {F.}~\bibnamefont {Fiore}}, \bibinfo
  {author} {\bibfnamefont {L.~A.}\ \bibnamefont {Antonelli}}, \bibinfo {author}
  {\bibfnamefont {S.}~\bibnamefont {Colafrancesco}}, \bibinfo {author}
  {\bibfnamefont {D.}~\bibnamefont {Guetta}}, \ and\ \bibinfo {author}
  {\bibfnamefont {A.}~\bibnamefont {Stamerra}},\ }\href {\doibase
  10.1051/0004-6361/201731452} {\bibfield  {journal} {\bibinfo  {journal}
  {Astron. Astrophys.}\ }\textbf {\bibinfo {volume} {607}},\ \bibinfo {pages}
  {A18} (\bibinfo {year} {2017})},\ \Eprint {http://arxiv.org/abs/1709.03497}
  {arXiv:1709.03497 [astro-ph.HE]} \BibitemShut {NoStop}%
\bibitem [{\citenamefont {Finke}\ \emph {et~al.}(2022)\citenamefont {Finke},
  \citenamefont {Ajello}, \citenamefont {Dominguez}, \citenamefont {Desai},
  \citenamefont {Hartmann}, \citenamefont {Paliya},\ and\ \citenamefont
  {Saldana-Lopez}}]{Finke:2022uvv}%
  \BibitemOpen
  \bibfield  {author} {\bibinfo {author} {\bibfnamefont {J.~D.}\ \bibnamefont
  {Finke}}, \bibinfo {author} {\bibfnamefont {M.}~\bibnamefont {Ajello}},
  \bibinfo {author} {\bibfnamefont {A.}~\bibnamefont {Dominguez}}, \bibinfo
  {author} {\bibfnamefont {A.}~\bibnamefont {Desai}}, \bibinfo {author}
  {\bibfnamefont {D.~H.}\ \bibnamefont {Hartmann}}, \bibinfo {author}
  {\bibfnamefont {V.~S.}\ \bibnamefont {Paliya}}, \ and\ \bibinfo {author}
  {\bibfnamefont {A.}~\bibnamefont {Saldana-Lopez}},\ }\href@noop {} {\
  (\bibinfo {year} {2022})},\ \Eprint {http://arxiv.org/abs/2210.01157}
  {arXiv:2210.01157 [astro-ph.GA]} \BibitemShut {NoStop}%
\bibitem [{\citenamefont {Owen}\ \emph {et~al.}(2022)\citenamefont {Owen},
  \citenamefont {Kong},\ and\ \citenamefont {Lee}}]{Owen2021}%
  \BibitemOpen
  \bibfield  {author} {\bibinfo {author} {\bibfnamefont {E.~R.}\ \bibnamefont
  {Owen}}, \bibinfo {author} {\bibfnamefont {A.~K.~H.}\ \bibnamefont {Kong}}, \
  and\ \bibinfo {author} {\bibfnamefont {K.-G.}\ \bibnamefont {Lee}},\
  }\href@noop {} {\bibfield  {journal} {\bibinfo  {journal} {Monthly Notices of
  the Royal Astronomical Society}\ }\textbf {\bibinfo {volume} {513}},\
  \bibinfo {pages} {2335} (\bibinfo {year} {2022})}\BibitemShut {NoStop}%
\bibitem [{\citenamefont {Cooray}(2016)}]{Cooray:2016jrk}%
  \BibitemOpen
  \bibfield  {author} {\bibinfo {author} {\bibfnamefont {A.}~\bibnamefont
  {Cooray}},\ }\href@noop {} {\bibfield  {journal} {\bibinfo  {journal} {Royal
  Society Open Science}\ }\textbf {\bibinfo {volume} {3}},\ \bibinfo {pages}
  {150555} (\bibinfo {year} {2016})},\ \Eprint
  {http://arxiv.org/abs/1602.03512} {arXiv:1602.03512 [astro-ph.CO]}
  \BibitemShut {NoStop}%
\bibitem [{\citenamefont {Singh}\ \emph {et~al.}(2021)\citenamefont {Singh},
  \citenamefont {Yadav},\ and\ \citenamefont {Meintjes}}]{Singh:2021yoi}%
  \BibitemOpen
  \bibfield  {author} {\bibinfo {author} {\bibfnamefont {K.~K.}\ \bibnamefont
  {Singh}}, \bibinfo {author} {\bibfnamefont {K.~K.}\ \bibnamefont {Yadav}}, \
  and\ \bibinfo {author} {\bibfnamefont {P.~J.}\ \bibnamefont {Meintjes}},\
  }\href {\doibase 10.1007/s10509-021-03957-z} {\bibfield  {journal} {\bibinfo
  {journal} {Astrophys. Space Sci.}\ }\textbf {\bibinfo {volume} {366}},\
  \bibinfo {pages} {51} (\bibinfo {year} {2021})},\ \Eprint
  {http://arxiv.org/abs/2105.14293} {arXiv:2105.14293 [astro-ph.CO]}
  \BibitemShut {NoStop}%
\bibitem [{\citenamefont {Yan-kun}\ and\ \citenamefont
  {Hou-dun}(2022)}]{Yan-kun:2022nmh}%
  \BibitemOpen
  \bibfield  {author} {\bibinfo {author} {\bibfnamefont {Q.}~\bibnamefont
  {Yan-kun}}\ and\ \bibinfo {author} {\bibfnamefont {Z.}~\bibnamefont
  {Hou-dun}},\ }\href {\doibase 10.1016/j.chinastron.2022.05.003} {\bibfield
  {journal} {\bibinfo  {journal} {Chin. Astron. Astrophys.}\ }\textbf {\bibinfo
  {volume} {46}},\ \bibinfo {pages} {42} (\bibinfo {year} {2022})}\BibitemShut
  {NoStop}%
\bibitem [{\citenamefont {Mattila}\ and\ \citenamefont
  {V\"ais\"anen}(2019)}]{Mattila:2019ybk}%
  \BibitemOpen
  \bibfield  {author} {\bibinfo {author} {\bibfnamefont {K.}~\bibnamefont
  {Mattila}}\ and\ \bibinfo {author} {\bibfnamefont {P.}~\bibnamefont
  {V\"ais\"anen}},\ }\href {\doibase 10.1080/00107514.2019.1586130} {\bibfield
  {journal} {\bibinfo  {journal} {Contemp. Phys.}\ }\textbf {\bibinfo {volume}
  {60}},\ \bibinfo {pages} {23} (\bibinfo {year} {2019})},\ \Eprint
  {http://arxiv.org/abs/1905.08825} {arXiv:1905.08825 [astro-ph.GA]}
  \BibitemShut {NoStop}%
\bibitem [{\citenamefont {Lunardini}\ \emph {et~al.}(2013)\citenamefont
  {Lunardini}, \citenamefont {Sabancilar},\ and\ \citenamefont
  {Yang}}]{Lunardini:2013iwa}%
  \BibitemOpen
  \bibfield  {author} {\bibinfo {author} {\bibfnamefont {C.}~\bibnamefont
  {Lunardini}}, \bibinfo {author} {\bibfnamefont {E.}~\bibnamefont
  {Sabancilar}}, \ and\ \bibinfo {author} {\bibfnamefont {L.}~\bibnamefont
  {Yang}},\ }\href {\doibase 10.1088/1475-7516/2013/08/014} {\bibfield
  {journal} {\bibinfo  {journal} {JCAP}\ }\textbf {\bibinfo {volume} {08}},\
  \bibinfo {pages} {014} (\bibinfo {year} {2013})},\ \Eprint
  {http://arxiv.org/abs/1306.1808} {arXiv:1306.1808 [astro-ph.HE]} \BibitemShut
  {NoStop}%
\bibitem [{\citenamefont {Caddy}\ \emph {et~al.}(2022)\citenamefont {Caddy},
  \citenamefont {Spitler},\ and\ \citenamefont {Ellis}}]{caddy2022towards}%
  \BibitemOpen
  \bibfield  {author} {\bibinfo {author} {\bibfnamefont {S.~E.}\ \bibnamefont
  {Caddy}}, \bibinfo {author} {\bibfnamefont {L.~R.}\ \bibnamefont {Spitler}},
  \ and\ \bibinfo {author} {\bibfnamefont {S.~C.}\ \bibnamefont {Ellis}},\
  }\href@noop {} {\bibfield  {journal} {\bibinfo  {journal} {arXiv preprint
  arXiv:2205.16002}\ } (\bibinfo {year} {2022})}\BibitemShut {NoStop}%
\bibitem [{\citenamefont {Bernal}\ \emph {et~al.}(2022)\citenamefont {Bernal},
  \citenamefont {Caputo}, \citenamefont {Sato-Polito}, \citenamefont
  {Mirocha},\ and\ \citenamefont {Kamionkowski}}]{Bernal:2022xyi}%
  \BibitemOpen
  \bibfield  {author} {\bibinfo {author} {\bibfnamefont {J.~L.}\ \bibnamefont
  {Bernal}}, \bibinfo {author} {\bibfnamefont {A.}~\bibnamefont {Caputo}},
  \bibinfo {author} {\bibfnamefont {G.}~\bibnamefont {Sato-Polito}}, \bibinfo
  {author} {\bibfnamefont {J.}~\bibnamefont {Mirocha}}, \ and\ \bibinfo
  {author} {\bibfnamefont {M.}~\bibnamefont {Kamionkowski}},\ }\href@noop {} {\
   (\bibinfo {year} {2022})},\ \Eprint {http://arxiv.org/abs/2208.13794}
  {arXiv:2208.13794 [astro-ph.CO]} \BibitemShut {NoStop}%
\bibitem [{\citenamefont {Kalashev}\ \emph {et~al.}(2019)\citenamefont
  {Kalashev}, \citenamefont {Kusenko},\ and\ \citenamefont
  {Vitagliano}}]{Kalashev:2018bra}%
  \BibitemOpen
  \bibfield  {author} {\bibinfo {author} {\bibfnamefont {O.~E.}\ \bibnamefont
  {Kalashev}}, \bibinfo {author} {\bibfnamefont {A.}~\bibnamefont {Kusenko}}, \
  and\ \bibinfo {author} {\bibfnamefont {E.}~\bibnamefont {Vitagliano}},\
  }\href {\doibase 10.1103/PhysRevD.99.023002} {\bibfield  {journal} {\bibinfo
  {journal} {Phys. Rev. D}\ }\textbf {\bibinfo {volume} {99}},\ \bibinfo
  {pages} {023002} (\bibinfo {year} {2019})},\ \Eprint
  {http://arxiv.org/abs/1808.05613} {arXiv:1808.05613 [hep-ph]} \BibitemShut
  {NoStop}%
\bibitem [{\citenamefont {Korochkin}\ \emph
  {et~al.}(2020{\natexlab{a}})\citenamefont {Korochkin}, \citenamefont
  {Neronov},\ and\ \citenamefont {Semikoz}}]{Korochkin:2019pzr}%
  \BibitemOpen
  \bibfield  {author} {\bibinfo {author} {\bibfnamefont {A.}~\bibnamefont
  {Korochkin}}, \bibinfo {author} {\bibfnamefont {A.}~\bibnamefont {Neronov}},
  \ and\ \bibinfo {author} {\bibfnamefont {D.}~\bibnamefont {Semikoz}},\ }\href
  {\doibase 10.1051/0004-6361/201936262} {\bibfield  {journal} {\bibinfo
  {journal} {Astron. Astrophys.}\ }\textbf {\bibinfo {volume} {633}},\ \bibinfo
  {pages} {A74} (\bibinfo {year} {2020}{\natexlab{a}})},\ \Eprint
  {http://arxiv.org/abs/1906.12168} {arXiv:1906.12168 [astro-ph.HE]}
  \BibitemShut {NoStop}%
\bibitem [{\citenamefont {Korochkin}\ \emph
  {et~al.}(2020{\natexlab{b}})\citenamefont {Korochkin}, \citenamefont
  {Neronov},\ and\ \citenamefont {Semikoz}}]{Korochkin:2019qpe}%
  \BibitemOpen
  \bibfield  {author} {\bibinfo {author} {\bibfnamefont {A.}~\bibnamefont
  {Korochkin}}, \bibinfo {author} {\bibfnamefont {A.}~\bibnamefont {Neronov}},
  \ and\ \bibinfo {author} {\bibfnamefont {D.}~\bibnamefont {Semikoz}},\ }\href
  {\doibase 10.1088/1475-7516/2020/03/064} {\bibfield  {journal} {\bibinfo
  {journal} {JCAP}\ }\textbf {\bibinfo {volume} {03}},\ \bibinfo {pages} {064}
  (\bibinfo {year} {2020}{\natexlab{b}})},\ \Eprint
  {http://arxiv.org/abs/1911.13291} {arXiv:1911.13291 [hep-ph]} \BibitemShut
  {NoStop}%
\bibitem [{\citenamefont {D'Olivo}\ \emph {et~al.}(2006)\citenamefont
  {D'Olivo}, \citenamefont {Nellen}, \citenamefont {Sahu},\ and\ \citenamefont
  {Van~Elewyck}}]{DOlivo:2005edp}%
  \BibitemOpen
  \bibfield  {author} {\bibinfo {author} {\bibfnamefont {J.~C.}\ \bibnamefont
  {D'Olivo}}, \bibinfo {author} {\bibfnamefont {L.}~\bibnamefont {Nellen}},
  \bibinfo {author} {\bibfnamefont {S.}~\bibnamefont {Sahu}}, \ and\ \bibinfo
  {author} {\bibfnamefont {V.}~\bibnamefont {Van~Elewyck}},\ }\href {\doibase
  10.1016/j.astropartphys.2005.11.005} {\bibfield  {journal} {\bibinfo
  {journal} {Astropart. Phys.}\ }\textbf {\bibinfo {volume} {25}},\ \bibinfo
  {pages} {47} (\bibinfo {year} {2006})},\ \Eprint
  {http://arxiv.org/abs/astro-ph/0507333} {arXiv:astro-ph/0507333} \BibitemShut
  {NoStop}%
\bibitem [{\citenamefont {Ruffini}\ \emph {et~al.}(2016)\citenamefont
  {Ruffini}, \citenamefont {Vereshchagin},\ and\ \citenamefont
  {Xue}}]{Ruffini:2015oha}%
  \BibitemOpen
  \bibfield  {author} {\bibinfo {author} {\bibfnamefont {R.}~\bibnamefont
  {Ruffini}}, \bibinfo {author} {\bibfnamefont {G.~V.}\ \bibnamefont
  {Vereshchagin}}, \ and\ \bibinfo {author} {\bibfnamefont {S.~S.}\
  \bibnamefont {Xue}},\ }\href {\doibase 10.1007/s10509-016-2668-5} {\bibfield
  {journal} {\bibinfo  {journal} {Astrophys. Space Sci.}\ }\textbf {\bibinfo
  {volume} {361}},\ \bibinfo {pages} {82} (\bibinfo {year} {2016})},\ \Eprint
  {http://arxiv.org/abs/1503.07749} {arXiv:1503.07749 [astro-ph.HE]}
  \BibitemShut {NoStop}%
\bibitem [{\citenamefont {Franceschini}(2021)}]{Franceschini:2021wkr}%
  \BibitemOpen
  \bibfield  {author} {\bibinfo {author} {\bibfnamefont {A.}~\bibnamefont
  {Franceschini}},\ }\href {\doibase 10.3390/universe7050146} {\bibfield
  {journal} {\bibinfo  {journal} {Universe}\ }\textbf {\bibinfo {volume} {7}},\
  \bibinfo {pages} {146} (\bibinfo {year} {2021})}\BibitemShut {NoStop}%
\bibitem [{\citenamefont {Nikishov}(1961)}]{nikishov1961absorption}%
  \BibitemOpen
  \bibfield  {author} {\bibinfo {author} {\bibfnamefont {A.}~\bibnamefont
  {Nikishov}},\ }\href@noop {} {\bibfield  {journal} {\bibinfo  {journal}
  {Zhur. Eksptl'. i Teoret. Fiz.}\ }\textbf {\bibinfo {volume} {41}} (\bibinfo
  {year} {1961})}\BibitemShut {NoStop}%
\bibitem [{\citenamefont {Fazio}\ and\ \citenamefont
  {Stecker}(1970)}]{Fazio:1970pr}%
  \BibitemOpen
  \bibfield  {author} {\bibinfo {author} {\bibfnamefont {G.~G.}\ \bibnamefont
  {Fazio}}\ and\ \bibinfo {author} {\bibfnamefont {F.~W.}\ \bibnamefont
  {Stecker}},\ }\href {\doibase 10.1038/226135a0} {\bibfield  {journal}
  {\bibinfo  {journal} {Nature}\ }\textbf {\bibinfo {volume} {226}},\ \bibinfo
  {pages} {135} (\bibinfo {year} {1970})}\BibitemShut {NoStop}%
\bibitem [{\citenamefont {Loeb}(2022)}]{Loeb:2022pdm}%
  \BibitemOpen
  \bibfield  {author} {\bibinfo {author} {\bibfnamefont {A.}~\bibnamefont
  {Loeb}},\ }\href {\doibase 10.3847/2515-5172/ac81d5} {\bibfield  {journal}
  {\bibinfo  {journal} {Res. Notes AAS}\ }\textbf {\bibinfo {volume} {6}},\
  \bibinfo {pages} {148} (\bibinfo {year} {2022})},\ \Eprint
  {http://arxiv.org/abs/2207.00671} {arXiv:2207.00671 [hep-ph]} \BibitemShut
  {NoStop}%
\bibitem [{\citenamefont {Breit}\ and\ \citenamefont
  {Wheeler}(1934)}]{Breit:1934zz}%
  \BibitemOpen
  \bibfield  {author} {\bibinfo {author} {\bibfnamefont {G.}~\bibnamefont
  {Breit}}\ and\ \bibinfo {author} {\bibfnamefont {J.~A.}\ \bibnamefont
  {Wheeler}},\ }\href {\doibase 10.1103/PhysRev.46.1087} {\bibfield  {journal}
  {\bibinfo  {journal} {Phys. Rev.}\ }\textbf {\bibinfo {volume} {46}},\
  \bibinfo {pages} {1087} (\bibinfo {year} {1934})}\BibitemShut {NoStop}%
\bibitem [{\citenamefont {Gould}\ and\ \citenamefont
  {Schr\'eder}(1967)}]{PhysRev.155.1404}%
  \BibitemOpen
  \bibfield  {author} {\bibinfo {author} {\bibfnamefont {R.~J.}\ \bibnamefont
  {Gould}}\ and\ \bibinfo {author} {\bibfnamefont {G.~P.}\ \bibnamefont
  {Schr\'eder}},\ }\href {\doibase 10.1103/PhysRev.155.1404} {\bibfield
  {journal} {\bibinfo  {journal} {Phys. Rev.}\ }\textbf {\bibinfo {volume}
  {155}},\ \bibinfo {pages} {1404} (\bibinfo {year} {1967})}\BibitemShut
  {NoStop}%
\bibitem [{\citenamefont {Ruffini}\ \emph {et~al.}(2010)\citenamefont
  {Ruffini}, \citenamefont {Vereshchagin},\ and\ \citenamefont
  {Xue}}]{Ruffini:2009hg}%
  \BibitemOpen
  \bibfield  {author} {\bibinfo {author} {\bibfnamefont {R.}~\bibnamefont
  {Ruffini}}, \bibinfo {author} {\bibfnamefont {G.}~\bibnamefont
  {Vereshchagin}}, \ and\ \bibinfo {author} {\bibfnamefont {S.-S.}\
  \bibnamefont {Xue}},\ }\href {\doibase 10.1016/j.physrep.2009.10.004}
  {\bibfield  {journal} {\bibinfo  {journal} {Phys. Rept.}\ }\textbf {\bibinfo
  {volume} {487}},\ \bibinfo {pages} {1} (\bibinfo {year} {2010})},\ \Eprint
  {http://arxiv.org/abs/0910.0974} {arXiv:0910.0974 [astro-ph.HE]} \BibitemShut
  {NoStop}%
\bibitem [{\citenamefont {Brown}\ \emph {et~al.}(1973)\citenamefont {Brown},
  \citenamefont {Mikaelian},\ and\ \citenamefont
  {Gould}}]{brown1973absorption}%
  \BibitemOpen
  \bibfield  {author} {\bibinfo {author} {\bibfnamefont {R.}~\bibnamefont
  {Brown}}, \bibinfo {author} {\bibfnamefont {K.}~\bibnamefont {Mikaelian}}, \
  and\ \bibinfo {author} {\bibfnamefont {R.}~\bibnamefont {Gould}},\
  }\href@noop {} {\bibfield  {journal} {\bibinfo  {journal} {Astrophysical
  Letters}\ }\textbf {\bibinfo {volume} {14}},\ \bibinfo {pages} {203}
  (\bibinfo {year} {1973})}\BibitemShut {NoStop}%
\bibitem [{\citenamefont {Zyla}\ \emph {et~al.}(2020)\citenamefont {Zyla} \emph
  {et~al.}}]{ParticleDataGroup:2020ssz}%
  \BibitemOpen
  \bibfield  {author} {\bibinfo {author} {\bibfnamefont {P.~A.}\ \bibnamefont
  {Zyla}} \emph {et~al.} (\bibinfo {collaboration} {Particle Data Group}),\
  }\href {\doibase 10.1093/ptep/ptaa104} {\bibfield  {journal} {\bibinfo
  {journal} {PTEP}\ }\textbf {\bibinfo {volume} {2020}},\ \bibinfo {pages}
  {083C01} (\bibinfo {year} {2020})}\BibitemShut {NoStop}%
\bibitem [{\citenamefont {Roulet}(1993)}]{Roulet:1992pz}%
  \BibitemOpen
  \bibfield  {author} {\bibinfo {author} {\bibfnamefont {E.}~\bibnamefont
  {Roulet}},\ }\href {\doibase 10.1103/PhysRevD.47.5247} {\bibfield  {journal}
  {\bibinfo  {journal} {Phys. Rev. D}\ }\textbf {\bibinfo {volume} {47}},\
  \bibinfo {pages} {5247} (\bibinfo {year} {1993})}\BibitemShut {NoStop}%
\bibitem [{\citenamefont {Barenboim}\ \emph {et~al.}(2005)\citenamefont
  {Barenboim}, \citenamefont {Mena~Requejo},\ and\ \citenamefont
  {Quigg}}]{Barenboim:2004di}%
  \BibitemOpen
  \bibfield  {author} {\bibinfo {author} {\bibfnamefont {G.}~\bibnamefont
  {Barenboim}}, \bibinfo {author} {\bibfnamefont {O.}~\bibnamefont
  {Mena~Requejo}}, \ and\ \bibinfo {author} {\bibfnamefont {C.}~\bibnamefont
  {Quigg}},\ }\href {\doibase 10.1103/PhysRevD.71.083002} {\bibfield  {journal}
  {\bibinfo  {journal} {Phys. Rev. D}\ }\textbf {\bibinfo {volume} {71}},\
  \bibinfo {pages} {083002} (\bibinfo {year} {2005})},\ \Eprint
  {http://arxiv.org/abs/hep-ph/0412122} {arXiv:hep-ph/0412122} \BibitemShut
  {NoStop}%
\bibitem [{\citenamefont {Bottcher}\ and\ \citenamefont
  {Dermer}(2005)}]{Bottcher:2005pj}%
  \BibitemOpen
  \bibfield  {author} {\bibinfo {author} {\bibfnamefont {M.}~\bibnamefont
  {Bottcher}}\ and\ \bibinfo {author} {\bibfnamefont {C.~D.}\ \bibnamefont
  {Dermer}},\ }\href {\doibase 10.1086/498615} {\bibfield  {journal} {\bibinfo
  {journal} {Astrophys. J. Lett.}\ }\textbf {\bibinfo {volume} {634}},\
  \bibinfo {pages} {L81} (\bibinfo {year} {2005})},\ \Eprint
  {http://arxiv.org/abs/astro-ph/0508359} {arXiv:astro-ph/0508359} \BibitemShut
  {NoStop}%
\bibitem [{\citenamefont {Finke}\ and\ \citenamefont
  {Patel}(2024)}]{Finke:2024ada}%
  \BibitemOpen
  \bibfield  {author} {\bibinfo {author} {\bibfnamefont {J.~D.}\ \bibnamefont
  {Finke}}\ and\ \bibinfo {author} {\bibfnamefont {P.}~\bibnamefont {Patel}},\
  }\href {\doibase 10.3847/1538-4357/ad3212} {\bibfield  {journal} {\bibinfo
  {journal} {Astrophys. J.}\ }\textbf {\bibinfo {volume} {965}},\ \bibinfo
  {pages} {44} (\bibinfo {year} {2024})},\ \Eprint
  {http://arxiv.org/abs/2403.07063} {arXiv:2403.07063 [astro-ph.HE]}
  \BibitemShut {NoStop}%
\bibitem [{\citenamefont {Lide}(2007)}]{lide2007crc}%
  \BibitemOpen
  \bibfield  {author} {\bibinfo {author} {\bibfnamefont {D.}~\bibnamefont
  {Lide}},\ }\href {https://books.google.fr/books?id=zIZGnwEACAAJ} {\emph
  {\bibinfo {title} {CRC Handbook of Chemistry and Physics, 88th Edition}}}\
  (\bibinfo  {publisher} {Taylor \& Francis},\ \bibinfo {year}
  {2007})\BibitemShut {NoStop}%
\bibitem [{\citenamefont {Abdo}\ \emph {et~al.}(2011)\citenamefont {Abdo} \emph
  {et~al.}}]{Fermi-LAT:2011nwz}%
  \BibitemOpen
  \bibfield  {author} {\bibinfo {author} {\bibfnamefont {A.~A.}\ \bibnamefont
  {Abdo}} \emph {et~al.} (\bibinfo {collaboration} {Fermi-LAT}),\ }\href
  {\doibase 10.1088/0004-637X/734/2/116} {\bibfield  {journal} {\bibinfo
  {journal} {Astrophys. J.}\ }\textbf {\bibinfo {volume} {734}},\ \bibinfo
  {pages} {116} (\bibinfo {year} {2011})},\ \Eprint
  {http://arxiv.org/abs/1104.2093} {arXiv:1104.2093 [astro-ph.HE]} \BibitemShut
  {NoStop}%
\bibitem [{\citenamefont {Ianni}(2014)}]{Ianni:2014bqa}%
  \BibitemOpen
  \bibfield  {author} {\bibinfo {author} {\bibfnamefont {A.}~\bibnamefont
  {Ianni}},\ }\href {\doibase 10.1016/j.dark.2014.06.002} {\bibfield  {journal}
  {\bibinfo  {journal} {Phys. Dark Univ.}\ }\textbf {\bibinfo {volume} {4}},\
  \bibinfo {pages} {44} (\bibinfo {year} {2014})}\BibitemShut {NoStop}%
\bibitem [{\citenamefont {Ackermann}\ \emph {et~al.}(2015)\citenamefont
  {Ackermann} \emph {et~al.}}]{Fermi-LAT:2014ryh}%
  \BibitemOpen
  \bibfield  {author} {\bibinfo {author} {\bibfnamefont {M.}~\bibnamefont
  {Ackermann}} \emph {et~al.} (\bibinfo {collaboration} {Fermi-LAT}),\ }\href
  {\doibase 10.1088/0004-637X/799/1/86} {\bibfield  {journal} {\bibinfo
  {journal} {Astrophys. J.}\ }\textbf {\bibinfo {volume} {799}},\ \bibinfo
  {pages} {86} (\bibinfo {year} {2015})},\ \Eprint
  {http://arxiv.org/abs/1410.3696} {arXiv:1410.3696 [astro-ph.HE]} \BibitemShut
  {NoStop}%
\bibitem [{\citenamefont {Maier}\ \emph {et~al.}(2020)\citenamefont {Maier},
  \citenamefont {Arrabito}, \citenamefont {Bernl\"ohr}, \citenamefont
  {Bregeon}, \citenamefont {Cumani}, \citenamefont {Hassan}, \citenamefont
  {Hinton},\ and\ \citenamefont {Moralejo}}]{Maier:2019afm}%
  \BibitemOpen
  \bibfield  {author} {\bibinfo {author} {\bibfnamefont {G.}~\bibnamefont
  {Maier}}, \bibinfo {author} {\bibfnamefont {L.}~\bibnamefont {Arrabito}},
  \bibinfo {author} {\bibfnamefont {K.}~\bibnamefont {Bernl\"ohr}}, \bibinfo
  {author} {\bibfnamefont {J.}~\bibnamefont {Bregeon}}, \bibinfo {author}
  {\bibfnamefont {P.}~\bibnamefont {Cumani}}, \bibinfo {author} {\bibfnamefont
  {T.}~\bibnamefont {Hassan}}, \bibinfo {author} {\bibfnamefont
  {J.}~\bibnamefont {Hinton}}, \ and\ \bibinfo {author} {\bibfnamefont
  {A.}~\bibnamefont {Moralejo}} (\bibinfo {collaboration} {CTA Consortium}),\
  }\href {\doibase 10.22323/1.358.0733} {\bibfield  {journal} {\bibinfo
  {journal} {PoS}\ }\textbf {\bibinfo {volume} {ICRC2019}},\ \bibinfo {pages}
  {733} (\bibinfo {year} {2020})},\ \Eprint {http://arxiv.org/abs/1907.08171}
  {arXiv:1907.08171 [astro-ph.IM]} \BibitemShut {NoStop}%
\bibitem [{\citenamefont {Abeysekara}\ \emph {et~al.}(2017)\citenamefont
  {Abeysekara} \emph {et~al.}}]{Abeysekara:2017mjj}%
  \BibitemOpen
  \bibfield  {author} {\bibinfo {author} {\bibfnamefont {A.~U.}\ \bibnamefont
  {Abeysekara}} \emph {et~al.},\ }\href {\doibase 10.3847/1538-4357/aa7555}
  {\bibfield  {journal} {\bibinfo  {journal} {Astrophys. J.}\ }\textbf
  {\bibinfo {volume} {843}},\ \bibinfo {pages} {39} (\bibinfo {year} {2017})},\
  \Eprint {http://arxiv.org/abs/1701.01778} {arXiv:1701.01778 [astro-ph.HE]}
  \BibitemShut {NoStop}%
\bibitem [{\citenamefont {Abeysekara}\ \emph
  {et~al.}(2019{\natexlab{a}})\citenamefont {Abeysekara} \emph
  {et~al.}}]{HAWC:2019xhp}%
  \BibitemOpen
  \bibfield  {author} {\bibinfo {author} {\bibfnamefont {A.~U.}\ \bibnamefont
  {Abeysekara}} \emph {et~al.} (\bibinfo {collaboration} {HAWC}),\ }\href
  {\doibase 10.3847/1538-4357/ab2f7d} {\bibfield  {journal} {\bibinfo
  {journal} {Astrophys. J.}\ }\textbf {\bibinfo {volume} {881}},\ \bibinfo
  {pages} {134} (\bibinfo {year} {2019}{\natexlab{a}})},\ \Eprint
  {http://arxiv.org/abs/1905.12518} {arXiv:1905.12518 [astro-ph.HE]}
  \BibitemShut {NoStop}%
\bibitem [{\citenamefont {Cao}\ \emph {et~al.}(2023{\natexlab{a}})\citenamefont
  {Cao} \emph {et~al.}}]{LHAASO:2023gne}%
  \BibitemOpen
  \bibfield  {author} {\bibinfo {author} {\bibfnamefont {Z.}~\bibnamefont
  {Cao}} \emph {et~al.} (\bibinfo {collaboration} {LHAASO}),\ }\href@noop {} {\
   (\bibinfo {year} {2023}{\natexlab{a}})},\ \Eprint
  {http://arxiv.org/abs/2305.05372} {arXiv:2305.05372 [astro-ph.HE]}
  \BibitemShut {NoStop}%
\bibitem [{\citenamefont {Cao}\ \emph {et~al.}(2023{\natexlab{b}})\citenamefont
  {Cao} \emph {et~al.}}]{LHAASO:2023rpg}%
  \BibitemOpen
  \bibfield  {author} {\bibinfo {author} {\bibfnamefont {Z.}~\bibnamefont
  {Cao}} \emph {et~al.} (\bibinfo {collaboration} {LHAASO}),\ }\href@noop {} {\
   (\bibinfo {year} {2023}{\natexlab{b}})},\ \Eprint
  {http://arxiv.org/abs/2305.17030} {arXiv:2305.17030 [astro-ph.HE]}
  \BibitemShut {NoStop}%
\bibitem [{\citenamefont {Abeysekara}\ \emph
  {et~al.}(2019{\natexlab{b}})\citenamefont {Abeysekara} \emph
  {et~al.}}]{HAWC:2018wju}%
  \BibitemOpen
  \bibfield  {author} {\bibinfo {author} {\bibfnamefont {A.~U.}\ \bibnamefont
  {Abeysekara}} \emph {et~al.} (\bibinfo {collaboration} {HAWC, IceCube}),\
  }\href {\doibase 10.3847/1538-4357/aaf5cc} {\bibfield  {journal} {\bibinfo
  {journal} {Astrophys. J.}\ }\textbf {\bibinfo {volume} {871}},\ \bibinfo
  {pages} {96} (\bibinfo {year} {2019}{\natexlab{b}})},\ \Eprint
  {http://arxiv.org/abs/1812.05682} {arXiv:1812.05682 [astro-ph.HE]}
  \BibitemShut {NoStop}%
\bibitem [{\citenamefont {Abeysekara}\ \emph {et~al.}(2018)\citenamefont
  {Abeysekara} \emph {et~al.}}]{Abeysekara:2018qho}%
  \BibitemOpen
  \bibfield  {author} {\bibinfo {author} {\bibfnamefont {A.~U.}\ \bibnamefont
  {Abeysekara}} \emph {et~al.},\ }\href {\doibase 10.3847/1538-4357/aad90c}
  {\bibfield  {journal} {\bibinfo  {journal} {Astrophys. J.}\ }\textbf
  {\bibinfo {volume} {865}},\ \bibinfo {pages} {57} (\bibinfo {year} {2018})},\
  \Eprint {http://arxiv.org/abs/1805.01847} {arXiv:1805.01847 [astro-ph.HE]}
  \BibitemShut {NoStop}%
\bibitem [{\citenamefont {Kogut}\ \emph {et~al.}(1993)\citenamefont {Kogut}
  \emph {et~al.}}]{Kogut:1993ag}%
  \BibitemOpen
  \bibfield  {author} {\bibinfo {author} {\bibfnamefont {A.}~\bibnamefont
  {Kogut}} \emph {et~al.},\ }\href {\doibase 10.1086/173453} {\bibfield
  {journal} {\bibinfo  {journal} {Astrophys. J.}\ }\textbf {\bibinfo {volume}
  {419}},\ \bibinfo {pages} {1} (\bibinfo {year} {1993})},\ \Eprint
  {http://arxiv.org/abs/astro-ph/9312056} {arXiv:astro-ph/9312056} \BibitemShut
  {NoStop}%
\bibitem [{\citenamefont {Wiedenbeck}\ \emph {et~al.}(2018)\citenamefont
  {Wiedenbeck} \emph {et~al.}}]{Wiedenbeck:2017ppg}%
  \BibitemOpen
  \bibfield  {author} {\bibinfo {author} {\bibfnamefont {M.~E.}\ \bibnamefont
  {Wiedenbeck}} \emph {et~al.},\ }\href {\doibase 10.22323/1.301.0016}
  {\bibfield  {journal} {\bibinfo  {journal} {PoS}\ }\textbf {\bibinfo {volume}
  {ICRC2017}},\ \bibinfo {pages} {016} (\bibinfo {year} {2018})}\BibitemShut
  {NoStop}%
\bibitem [{\citenamefont {Aartsen}\ \emph {et~al.}(2018)\citenamefont {Aartsen}
  \emph {et~al.}}]{IceCube:2018fhm}%
  \BibitemOpen
  \bibfield  {author} {\bibinfo {author} {\bibfnamefont {M.~G.}\ \bibnamefont
  {Aartsen}} \emph {et~al.} (\bibinfo {collaboration} {IceCube}),\ }\href
  {\doibase 10.1103/PhysRevD.98.062003} {\bibfield  {journal} {\bibinfo
  {journal} {Phys. Rev. D}\ }\textbf {\bibinfo {volume} {98}},\ \bibinfo
  {pages} {062003} (\bibinfo {year} {2018})},\ \Eprint
  {http://arxiv.org/abs/1807.01820} {arXiv:1807.01820 [astro-ph.HE]}
  \BibitemShut {NoStop}%
\end{thebibliography}%

\end{document}